\documentstyle[preprint,aps]{revtex}
\begin{document}
\title{Persistent current of two-chain Hubbard model with impurities}
\author{Hiroyuki Mori}
\address{Department of Materials Science,
Hiroshima University,\\
Higashi-Hiroshima 739-8526, Japan}
\maketitle
%
%
\begin{abstract}
The interplay between impurities and interactions is studied
in the gapless phase of two-chain Hubbard model in order
to see how the screening of impurity potentials due to 
repulsive interactions in single-chain model
will be changed by increasing the number of channels.
Renormalization group calculations show that charge
stiffness, and hence persistent current, of the two-chain
model are less enhanced by interactions
than single chain case.
\end{abstract}
\tighten
%
%
\section{Introduction}
Interacting electrons in a disordered system
constitue one of the most challenging problems. 
Questions on roles of interactions and disorder
in quantum wires were again cast by the experiments
of persistent current in mesoscopic rings\cite{levy}.
The ground state of a ring system with a magnetic flux
breaks time reversal symmerty and has a finite 
circulating current as a periodic function of the flux.
This persistent current was predicted theoretically
long time ago\cite{byers},
but we had to wait a few decades until technological progress 
allowed to observe the current.
The experiments revealed that the ring supports a
large current even with modest amount of impurities.
Non-interacting electrons with impurities are shown
\cite{cheung}
to have persistent current with a factor $l/L$ multiplied,
where $l$ is the elastic mean free path and $L$ the ring size.
The current observed in the experiments
was much larger than this value
and other effects had to be taken into account to
explain the large current. Electron-electron interactions were
considered to be the first candidate as an origin of
the large current, because a simple consideration
gives a scenario that repulsive interactions would
prevent particles from gathering in deep impurity
holes and hence the interacting electrons would not be
easily pinned by impurities, compared to non-interacting
electrons. However the problem is not that simple
as shown in the studies of one-dimensional (1D) system.
For example, the roles of interactions are quite different
between the systems with spinning and spinless electrons.
The detail of the difference is summarized in section II.

Since 1D systems allow us to study the physical properties
rather rigorously with the help of bosonization and
renormalization group (RG) techniques. 
Approaching the problem of persistent current from a viewpoint
of 1D systems, hence, has this kind of advantage.
Analytical and numerical studies have been done
for interacting electrons in a single chain with disorder
potentials. Summary of their results is given also
in section II.

In order to approach more realistic model, one need to
consider multi-channel effects on the chain.
Recent growing interests in few-coupled chains
are mostly focused on pure systems and there have been
only a few studies on ladder electron systems with
impurity potentials. In this paper we study two-chain
Hubbard model with impurities as a first step
from a single chain towards a finite cross section ring.
In section III we derive RG equations of
the renormalized strengths
of impurity potentials in the gapless phase, and show
how the relation between interactions and disorder is changed
by increasing the number of channels. Summary is given
in section IV.
%
%
\section{Single chain}
Let us first begin by summarizing the impurity-interaction
effects on persistent current of single chains.
For spinless electron models with impurities,
analytical and numerical studies
showed that persistent current is suppressed by interparticle
repulsive correlation as far as
the impurity potentials are
weak enough\cite{giamarchi2,bouzerar,mori}.
A RG calculation gives the effective backward
impurity scattering $W$ is renormalized as
${\rm d}W/{\rm d}l=(3-2K_{\rho})W$
where $K_{\rho}$ is the Luttinger parameter. \cite{giamarchi2}
The larger repulsive interactions
give rise to the smaller $K_{\rho}$
and therefore to the stronger effective impurity potentials,
which would diminish the persistent current.
This is due to the enhancement of charge density wave (CDW)
correlations in the ground state of the system.
The repulsive interactions would enhance
the CDW correlations, which make the system easily pinned by the
impurities therefore make the persistent current smaller
than the value of non-interacting systems.

In the presence of the strong impurity potentials,
on the other hand, the wave functions are almost localized.
When the interactions are turned on between the localized
particles, they start to escape from each other and
the localization of the particles in the deep impurity potentials
is relaxed effectively. As a consequence,
the interactions tend to
enhance the persistent current of systems with
strong impurity potentials\cite{mori}.

In the experimental point of view, the above results are
disappointing because the materials used in the experiments
are all in the diffusive or ballistic regime so the scenario
of strong impurity potentials cannot be applied.

The situation changes drastically, however,
when the spin degrees
of freedom are taken into account\cite{giamarchi1}.
The renormalized impurity potential is given by
${\rm d}W/{\rm d}l=(3-K_{\rho}-K_{\sigma}-y)W$,
where $K_{\nu}\ \ (\nu=\rho, \sigma)$ is the Luttinger parameters
of charge and spin sectors respectively and $y$ measures
the backward scattering strength between electrons of opposite
spins. For Hubbard model with small $U$, this RG equation becomes
\begin{equation}
{\rm d}W/{\rm d}l=(1-U/\pi v_{F})W,
\label{W}
\end{equation}
where $v_F$ is the Fermi velocity.
Here electron-electron interactions would screen impurity potentials.
Since spin density wave (SDW) correlation is dominant
in the ground state of the repulsive
Hubbard model, the interaction $U$ makes
the particle density uniform,
and therefore make the coupling of the density to the impurities
weak. This role of interactions is a good news to explain
the experimental data: electron-electron interactions weaken
the impurity effects and the persistent current
would be enhanced from the value of non-interacting systems.

Note that, while $2k_F$ component of density-impurity coupling
is renormalized to smaller value due to the interactions as
shown, $4k_F$ component grows in the presence of the interactions.
Since $k_F$ of spinning electrons is equal to
$2k_F$ of spinless fermions, $4k_F$ coupling in infinite
$U$ Hubbard model corresponds to $2k_F$ coupling
of spinless model where the $2k_F$ coupling is enhanced 
by the interactions as stated above. Therefore as $U$ increases
the effective strength of impurity potentials decreases
as long as $U$ is small and then increases for large $U$
because of the effect of $4k_F$ component\cite{morihamada}.

Those stories are about short range interactions.
Then, what about
long-range Coulomb interactions? Since it is shown\cite{schulz1}
that long-range interactions make $4k_F$ density
correlations dominant, the interaction would enhance
the effective strengh of impurity potentials and hence
suppress persistent current.
The suppression of persistent current by Coulomb interactions
was observed in a numerical calculation\cite{kato}.

Those analyses have almost revealed the interplay
between interactions and impurities in single
chain systems. For more realistic discussions, however, one
has to consider multi-channel effects because the real ring systems
have finite cross sections. Although it is shown
that the persistent current of single-chain
Hubbard model would be
enhanced by the presence of electron-electron interactions, it is
not clear whether this effect still remains in multi-channel
systems and, more precisely, whether this effect will
become stronger or weaker as the number of channels
increases. In order to see this, we study two-channel
Hubbard model as a starting point toward multi-channel
models.
%
%
\section{Two-chain ladder}
In this section we study two-chain Hubbard model which
is described by $H=H_0+H_{int}$;
\begin{eqnarray}
H_0&=&-t\sum_{\alpha, s, i}c^{\dagger}_{\alpha s i}
c_{\alpha s i+1}-t_{\perp}\sum_{s, i}
c^{\dagger}_{1 s i}c_{2 s i}+\mbox{h.c.},\\
H_{int}&=&U\sum_{\alpha i}n_{\alpha \uparrow i}
n_{\alpha \downarrow i},
\end{eqnarray}
where $\alpha\ \ (=1, 2)$ is chain index.
$H_0$ can be diagonalized by $a_{o,\pi}=(c_1\pm c_2)/\sqrt{2}$,
where $o$ and $\pi$ denote the bonding
and anti-bonding bands respectively.
Using the standard
bosonization scheme for fermions with spins, we can describe
the system by four fields
$\phi^{\pi}_{\rho}, \phi^{\pi}_{\sigma}, \phi^{o}_{\rho},%
\phi^{o}_{\sigma}$.
Then, introducing a linear combination of the fields:
$\phi_{\nu\pm}=(\phi^{o}_{\nu}\pm\phi^{\pi}_{\nu})/\sqrt{2}$
where $\nu = \rho, \sigma$, we have the Hamiltonian,
$H=H_1+H_2+H_3$;
\begin{eqnarray}
H_1&=&\sum_{\parbox[t]{2cm}{$\nu=\rho,\sigma$\\$r=\pm$}}
\int\frac{{\rm d}x}{2\pi}[u_{\nu r}K_{\nu r}(\pi\Pi_{\nu r})^2
+\frac{u_{\nu r}}{K_{\nu r}}(\partial_x\phi_{\nu r})^2],
\label{H1}\\
H_2&=&\frac{1}
{2(\pi a)^2}\int{\rm d}x\{
g^{(2)}_{oo\pi\pi}\cos 2\theta_{\rho -}\cos 2\phi_{\sigma -}
-g^{(2)}_{o\pi o\pi}\cos 2\phi_{\rho -}\cos 2\theta_{\sigma -}
\label{H2}\\&&
+g^{(4)}_{oo\pi\pi}\cos 2\phi_{\rho -}\cos 2\theta_{\rho -}
-g^{(4)}_{o\pi o\pi}\cos 2\phi_{\sigma -}
\cos 2\theta_{\sigma -}\},
\nonumber\\
H_3&=&\frac{1}{2(\pi a)^2}\int{\rm d}x
\{g^{(1)}_{oooo}\cos 2
\phi_{\sigma -}
+g^{(1)}_{o\pi\pi o}
\cos 2\phi_{\rho -}
+g^{(1)}_{o\pi o\pi}
\cos 2\theta_{\sigma -}
-g^{(1)}_{oo\pi\pi}
\cos 2\theta_{\rho -}\}\cos 2\phi_{\sigma +},
\label{H3}
\end{eqnarray}
where $a$ is the lattice constant.
$H_2$ is associated with interband processes induced by
intrachain forward scattering, and $H_3$ with the intrachain
backward scattering.
The sign of each term in Eqs.(\ref{H2}) and (\ref{H3})
is determined according to a simple algebra of Majorana Fermions
\cite{schulz2},
introduced to preserve the proper anticommutation relations
between Fermion fields with differing band and spin indices.
The interaction parameter $g_{\alpha\beta\gamma\delta}$
represents the scattering from $(\delta, \gamma)$ to
$(\alpha, \beta)$. All $g$'s in Eqs. (\ref{H2}) and (\ref{H3})
are equal to $U$ for the case of Hubbard model.
We used in Eqs. (\ref{H2}) and (\ref{H3}) that
$g^{(1)}_{oooo}=g^{(1)}_{\pi\pi\pi\pi}$
and the similar properties of $g$'s.

The impurity scatterings are described by
\begin{equation}
H_{imp}=\sum_{
\alpha ,s,i}
V^{imp}_{\alpha i}n_{\alpha s i}.
\end{equation}
Applying the bosonization to $H_{imp}$ and separate it
into the forward and backward scattering parts,
$H_{imp}=H_f+H_b$, we get
\begin{eqnarray}
H_f&=&\int{\rm d}x\eta_1 (x)(\partial_x\phi_{\rho}^o
+\partial_x\phi_{\rho}^{\pi})\\\nonumber
&&+\int{\rm d}x\frac{\eta_2}{\pi a}
[e^{i(\phi_{\rho -}+\theta_{\rho -})}
\cos (\phi_{\sigma -}+\theta_{\sigma -})
+e^{i(-\phi_{\rho -}+\theta_{\rho -})}
\cos (\phi_{\sigma -}-\theta_{\sigma -})+\mbox{h.c}]
,\\
H_b&=&
\frac{1}{\pi\alpha}\int{\rm d}x\xi_1 (x)\{
e^{-i(\phi_{\rho +}+\phi_{\rho -})}\cos (\phi_{\sigma +}+
\phi_{\sigma -})+e^{-i(\phi_{\rho +}-\phi_{\rho -})}
\cos (\phi_{\sigma +}-\phi_{\sigma -})\}+\mbox{h.c.}
\nonumber\\
&&+\frac{1}{\pi\alpha}\int{\rm d}x\xi_2(x)\{
e^{-i(\phi_{\rho +}+\phi_{\sigma +})}\cos (\theta_{\rho -}+
\theta_{\sigma -})+e^{-i(\phi_{\rho +}-\phi_{\sigma +})}
\cos (\theta_{\rho -}-\theta_{\sigma -})\}+\mbox{h.c}.
\end{eqnarray}
$\eta_1$ and $\xi_1$ are the random potential fields
within $o$ and $\pi$  bands, and $\eta_2$ and $\xi_2$ are the random
hopping between $o$ and $\pi$ bands. 
In order to make a quantitative comparison with
the single chain case, we use the same
distribution for $\eta$ and $\xi$ as used in
the single chain model.
In the single chain model we assumed Gaussian distribution
and $\overline{\eta (0)\eta^(x)}=D\delta(x)$, 
$\overline{\xi^*(0)\xi (x)}=W\delta(x)$.
We assume the same for $V^{imp}_{\alpha i}$,
that is, we have
$\overline{\eta_i(0)\eta_j(x)}=(D_i/2)\delta_{ij}\delta(x)$ and
$\overline{\xi_i^*(0)\xi_j(x)}=(W_i/2)\delta_{ij}\delta(x)$,
where $i,j=1,2$.
The factor $1/2$ appeared in the transformation from
$c_{1,2}$ to $a_{o,\pi}$.

The Hamiltonian $H$ has been studied analytically
\cite{schulz2,fabrizio,balents,nagaosa} and numerically
\cite{noack,endres,asai,yamaji,kuroki}
and is shown to have a rich phase diagram. The model including
impurities, that is $H+H_{imp}$, was recently
investigated using a RG method\cite{orignac,fujimoto}.
It was found in a spin gap phase of the repulsive
Hubbard ladder that the dominant coupling between charge density
and impurity potentials is the $4k_F$ Fourier component,
which is already shown in the single chain models to be reduced
by the renormalizations. Therefore the persistent current
of the spin gap phase is suppressed by the repulsive interactions,
at variance with the single Hubbard chain.

The spin gap phase has attracted a lot of attentions in connection
with the possibility of superconductivity. But here our final
concern is to see the interplay between interactions and impurities
in multi-channel systems, keeping the experimentally relevant
situations in mind. Since the ground states are
gapless in the two limits,
a single channel and the real ring with many channels,
we would like to focus
on the {\em gapless} phase of the two-chain model \cite{sandler}
in order to start a systematic interpolation between the limits,
although the gapless phase occupies only a small area in
the phase diagram.\cite{balents}

Here we first study the effect of $H_b$ term.
Using the replica trick for the random variables and following
the standard RG procedure, we get the RG equations of the
backward impurity scatterings $W_1$ and $W_2$
in the gapless phase;
\begin{eqnarray}
\frac{{\rm d}W_1}{{\rm d}l}&=&
[3-(K_{\rho +}+K_{\rho -}+K_{\sigma +}
+K_{\sigma -})/2-
(g^{(1)}_{oooo}+g^{(1)}_{o\pi\pi o})/2\pi u_{\sigma}]
W_1,
\label{W1}\\
\frac{{\rm d}W_2}{{\rm d}l}&=&
[3-(K_{\rho +}+\frac{1}{K_{\rho -}}+K_{\sigma +}
+\frac{1}{K_{\sigma -}})/2-
(g^{(1)}_{o\pi o\pi}-g^{(1)}_{oo\pi\pi})/2\pi u_{\sigma}]
W_2.
\label{W2}
\end{eqnarray}

The renormalizations to $g^{(1)}$ in
Eqs. (\ref{W1}) and (\ref{W2}) are described by
\begin{eqnarray}
\frac{{\rm d}g^{(1)}_{oooo}}{{\rm d}l}&=&
(2-K_{\sigma +}-K_{\sigma -})g^{(1)}_{oooo}
-\frac{W_1a}{u_{\sigma}},\\
\frac{{\rm d}g^{(1)}_{o\pi\pi o}}{{\rm d}l}&=&
(2-K_{\sigma +}-K_{\rho -})g^{(1)}_{o\pi\pi o}
-\frac{W_1a}{u_{\sigma}},\\
\frac{{\rm d}g^{(1)}_{o\pi o\pi}}{{\rm d}l}&=&
(2-K_{\sigma +}-\frac{1}{K_{\rho -}})g^{(1)}_{o\pi o\pi}
-\frac{W_2a}{u_{\sigma}},\\
\frac{{\rm d}g^{(1)}_{oo\pi\pi}}{{\rm d}l}&=&
(2-K_{\sigma +}-\frac{1}{K_{\sigma -}})g^{(1)}_{oo\pi\pi}
+\frac{W_2a}{u_{\sigma}}.
\end{eqnarray}

In the case of Hubbard model with small $U$ and small disorder,
we ignore the renormalization
to $K$ and $g$ and then we get 
\begin{eqnarray}
\frac{{\rm d}W_1}{{\rm d}l}&=&
[1-U/\pi v_F]W_1,
\label{W12}\\
\frac{{\rm d}W_2}{{\rm d}l}&=&
[1+O((U/v_F)^2)]W_2
\label{W22}
\end{eqnarray}
The coefficient of $(U/v_F)^2$ term in Eq. (\ref{W22})
is positive.
First, we notice that Eq. (\ref{W12}) has the same
form as Eq. (\ref{W}), and the renormalization to
the backward impurity scatterings within the bands
is unchanged even when two chains are coupled,
which is to be contrasted to
the case of spin gap phase\cite{orignac}.
The impurity scatterings between the bands behave
differently, however. The presence of the electron-electron
interactions leads to the larger $W_2$, although
the dependence of $W_2$ on $U$ is rather weak
since the leading order in Eq. (\ref{W22})
is $(U/v_F)^2$.

The current coupled to the flux penetrating the ring,
is $\sum_s(j_{1s}+j_{2s})=\sum_s(j_{os}+j_{\pi s})=%
(2/\sqrt{\pi})\Pi_{\rho +}$. Therefore only
($\rho +$) mode contributes to persistent current
and charge stiffness $D$, where the latter is given by
$4K_{\rho +}u_{\rho +}$. Then the RG equation of
charge stiffness is
\begin{equation}
\frac{{\rm d}D}{{\rm d}l}=-\frac{2DK_{\rho +}u_{\rho +}
a}{\pi u_{\sigma +}^3}W_{+},
\label{D}
\end{equation}
where $W_{+}=(W_1+W_2)/2$.
Remember $W_1$ behaves the same way as in the single-chain case,
namely $U$ suppresses $W_1$, whereas $W_2$ increases with $U$.
The RG equation of $D$ for single Hubbard chain is
${\rm d}D/{\rm d}l=-(2DK_{\rho}u_{\rho}a/\pi u_{\sigma}^3)W$.
One then see $D$ of the gapless phase of
two chain model {\em is}
enhanced by the interaction $U$, which is in contrast
to the spin gap phase, but the enhancement is {\rm weaker}
than single chain.
Namely, the enhancement of charge
stiffness (therefore persistent current) due to the
electron-electron repulsive interactions becomes
smaller by doubling the number of channels.

The forward scatterings from impurities, $H_f$, were
already studied in Ref. \cite{orignac,fujimoto}.
The first term in $H_f$ can be absorbed in transforming
the definition of $\phi_{\rho\pm}$ and then has
no effect. The second term, however, was shown to make
the interaction terms effectively weak,
which results in that
the enhancement of the persistent current due to
the presence of the interactions would become
even smaller in two chains than in single chain.
%
%
\section{Summary}
We studied the impurity effect on the gapless
phase of the two-chain Hubbard model, as
a first step from a single chain analysis
toward the finite cross section ring.
In contrast to the spin gap phase,
the RG calculations show that the charge stiffness,
and hence persistent current, would be enhanced
by the repulsive interactions. The enhancement,
however, is smaller than in single Hubbard chain.
This indicates a possibility that the large persistent
current observed in the real rings with finite
cross sections might not be explained by
simply extending the mechanism of 
interaction-enhanced persistent current in single
Hubbard chain.
In order to draw more definite conclusion,
one has to check on three chain system and
work in this direction is in progress.
%
%

\end{document}